\documentclass{iopart}
\usepackage{graphicx}  
\usepackage{latexsym}

\jl{6}        
\eqnobysec    

\def\beq{\begin{equation}}
\def\eeq{\end{equation}}

\def\rmd{{\rm d}}

\begin{document}

\title[C metric: the equatorial plane and Fermi coordinates]
{C metric: the equatorial plane and Fermi coordinates}

\author{
Donato Bini$^* {}^\S{}^\P$,
Christian Cherubini$^\dagger {}^\S$,
Simonetta Filippi$^\dagger {}^\S$ and
Andrea Geralico${}^\S$
}

\address{
  ${}^*$\
Istituto per le Applicazioni del Calcolo ``M. Picone'', CNR I-00161 Rome, Italy
}
\address{
  ${}^\S$\
  International Center for Relativistic Astrophysics,
  University of Rome, I-00185 Rome, Italy
}
\address{
${}^\P$
  INFN - Sezione di Firenze, Polo Scientifico, Via Sansone 1, 
  I-50019, Sesto Fiorentino (FI), Italy 
}
\address{
${}^\dagger$\
Facolt\'a di Ingegneria, Universit\`a Campus Biomedico, Via E. Longoni 83,  I-00155 Roma, Italy
}

\begin{abstract}
We discuss geodesic motion in the vacuum C metric using Bondi-like spherical coordinates, with special attention to the role played by the \lq\lq equatorial plane."
We show that the spatial trajectory of photons on such a hypersurface is formally the same of photons on the equatorial plane of the Schwarzschild spacetime, apart from an energy shift involving the spacetime acceleration parameter.
Furthermore, we show that photons starting their motion from this hypersurface with vanishing component of the momentum along $\theta$, remain confined on it, differently from the case of massive particles. This effect is shown to have a counterpart also in the massless limit of the C metric, i.e. in Minkowski spacetime.
Finally, we give the explict map between Bondi-like spherical coordinates and Fermi coordinates (up to the second order) for the world line of an observer at rest at a fixed spatial point of the equatorial plane of the C metric, a  result which may be eventually useful to estimate both the mass and the acceleration parameter of accelerated sources.
\end{abstract}

\pacno{04.20.Cv}

\section{Vacuum C Metric}

The vacuum C metric was first discovered by Levi-Civita \cite{LC} in 1918 within a class of Petrov type D (degenerate) static vacuum metrics.
However, over the years it has been rediscovered many times: by Newman and Tamburino \cite{newtam} in 1961, by Robinson and Trautman \cite{robtra} in 1961 and again by Ehlers and Kundt \cite{ehlkun} ---who called it the C metric---
in 1962. It is worth to mention that a charged and spinning extension of the C metric exist: the former has been studied in detail by Kinnersley and Walker \cite{kin69,kinwal}, the latter instead has been deeply investigated by Bicak, Pravda, Pravdov\'a, Farhoosh and Zimmerman \cite{bicpra,prapra,Farh}. 
In general, the spacetime represented by the C metric contains one or, via an extension, two uniformly accelerated particles as explained in \cite{kinwal,bon83} and
a description 
of the geometric properties as well as the various extensions of the C metric is contained in \cite{ES}, which should be consulted for a more complete list of references.
The main property of the C metric is the existence of two hypersurface-orthogonal  Killing vectors, one of which is
timelike (showing the static property of the metric) in the spacetime region of interest in this work, as indicated below.

It is convenient to work with the Bondi-like spherical  coordinate system $x^0=u$, $x^1=r$, $x^2=\theta$, $x^3=\phi$, so that the C metric has  the form
\begin{equation}
\label{cmetu}
\rmd s^2= -H \rmd u^2 - 2 \rmd u \rmd r + 2A r^2 s\,\rmd u \rmd \theta +\frac{r^2s^2 }{G} \rmd \theta^2 +r^2 G \rmd \phi^2  ,
\end{equation} 
where $G$ and $H$ are given by
\begin{eqnarray}
\fl\quad
G(\theta)&=& s^2 -2mA c^3 , \nonumber \\
\fl\quad
H(r,\theta)&=&1-\frac{2m}{r}-A^2r^2 (s^2-2mAc^3)-2Arc(1+3mAc)+6mAc ,
\end{eqnarray}
and the simplified notation $c\equiv\cos\theta$ and $s\equiv\sin\theta$ has been introduced. 
The constants $M\ge 0$ and $A\ge 0$ denote the mass and acceleration of the source, respectively. The metric (\ref{cmetu}) is a nonlinear superposition of two metrics which are special cases, one  associated with the Schwarzschild black hole (the case $A=0$) and the other flat spacetime in uniformly accelerating coordinates (the case $M=0$) 
\cite{kin69,kinwal,Farh}.
Moreover, the C metric as written is assumed to have signature +2 with $\tilde F>0, \tilde G>0$, which limit the ranges of the nonignorable coordinates. To avoid a naked singularity, one must restrict the two parameters by the condition $MA<1/(3\sqrt{3})$  \cite{Farh,pavda,podol}.
Units here are choosen so that the speed of light is taken to be $1$ and hence the previous notation does not create any confusion.

\begin{figure}
\typeout{*** EPS figure 1}
\begin{center}
\includegraphics[scale=0.45]{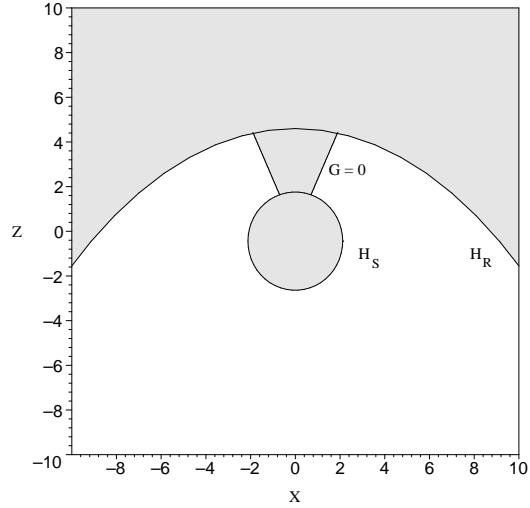}
\end{center}
\caption{ 
The accessible spacetime region (unshaded) in the polar coordinate representation of the $r$-$\theta$ plane ($\theta$ measured from the upward vertical $Z$-axis) is shown for $MA=0.1$. The upper boundary curve $H_R$ represents the Rindler horizon while the circle $H_S$ is the Schwarzschild horizon. The forbidded conical region corresponds to negative values of the metric function $G$, i.e., to signature changes which are not considered here. 
}
\label{fig:1}
\end{figure}

The C metric has event horizons (which are also Killing horizons) given by hypersurfaces of the form $r=r(\theta)$ that are solutions of $H=0$, and can be determined exactly \cite{bcmprd}. 
To study the location of the horizons it is useful to introduce an acceleration length scale based on $A>0$ given by $L_A= 1/(3\sqrt{3}A)$, as well as the quantities:
\beq
\fl\quad
U=-\frac{1}{\sqrt{3}}\cos \left(\frac13 {\rm arccos}\, \frac{M}{L_A} \right), \qquad 
V=\frac{1}{\sqrt{3}}\sin \left(\frac13 {\rm arccos}\, \frac{M}{L_A} \right).
\eeq
It turns out that there is an inner horizon which is a deformation of the Schwarschild one, located at
\beq
\label{horS}
r_S= \frac1A \frac{\sqrt{3}V-U}{[1+(\sqrt{3}V-U)c]},
\eeq
and an outer horizon which is a deformation of the Rindler one, located at
\beq
\label{horR}
r_R= \frac1A \frac{2U}{(1+2Uc)},
\eeq
while the third real radial root of $H=0$ corresponds to a negative value of $r$.
In the polar coordinate plane of the noningnorable coordinates $(r,\theta)$, depicted in terms of the auxiliary Cartesian-like coordinates $X=r\sin \theta$ (horizontal), $Z=r\cos\theta$ (vertical), the accessible region of spacetime is shown as the unshaded region in Fig.~\ref{fig:1} for $MA=0.1$, which satisfy the condition $MA<1/(3\sqrt{3})$.

 
The C metric also admits a conformal Killing tensor \cite{kin69,kinwal},
\beq
\label{ckt}
\fl\quad
P=r^2\left( H\rmd u^2+2 \rmd u \rmd r -2Asr^2 \rmd u \rmd \theta +\frac{r^2s^2}{G}\rmd \theta^2+r^2 G \rmd \phi^2 \right),
\eeq
symmetric and traceless, satisfying the equation $P_{(\alpha \beta ; \gamma)}-1/3 g_{(\alpha \beta}P^\delta{}_{\gamma) ; \delta}=0$
to which a conserved quantity for null geodesics (with four momentum $P^\alpha$)
\beq
Q=\frac12 P_{\alpha\beta}\, P^\alpha P^\beta,
\eeq
is associated.

A natural family of observers in the C metric is represented by the (static) timelike congruence of the coordinate time world lines.
Let us denote the observer 4-velocity by $e_{\hat u}=\frac{1}{M}\partial_u$,
and use the lapse-shift notation for the metric components:
\begin{eqnarray}
g_{uu}=-M^2, \quad g_{ua}=M^2M_a, \quad  g_{ab}=\gamma_{ab}+M^2M_aM_b, 
\end{eqnarray}
where $M$ is the laspe factor and $M_a dx^a$, $a=1,2,3$ is the sfift 1-form field;
the line element (\ref{cmetu}) then takes the form
\beq
\rmd s^2=-M^2(\rmd u-M_a\rmd x^a)^2+\gamma_{ab}\rmd x^a\rmd x^b,
\eeq 
where the functions $M,M_a$ and $\gamma_{ab}$ are identified by using Eq.~(\ref{cmetu}):
\begin{eqnarray}
\fl\quad  &M=\sqrt{H}>0, \quad M_r=-\frac{1}{H}<0, \quad M_\theta= \frac{Ar^2s}{H}>0,\quad M_\phi=0, \nonumber \\
\fl\quad  &\gamma_{rr}=1/H, \quad \gamma_{\theta\theta}=r^2s^2(1/G+A^2r^2/H), \quad \gamma_{r\theta}=-Ar^2s/H, \quad\gamma_{\phi\phi}=r^2G,
\end{eqnarray}
and all the other metric components vanish.
The dual of $e_{\hat u}$ is denoted by $\omega^{\hat u}=M(\rmd u -M_r \rmd r -M_\theta \rmd \theta)$ and
an orthonormal triad adapted to $e_{\hat u}$ is given by
\begin{eqnarray}
\fl\qquad
e_{\hat r}=\frac{1}{M}[-\frac{1}{M_r}\partial_r -\partial_u],\quad
e_{\hat \theta}= \frac{1}{\sqrt{g_{\theta\theta}}}[\partial_\theta -\frac{M_\theta}{M_r}\partial_r],\quad
e_{\hat \phi}= \frac{1}{\sqrt{g_{\phi\phi}}}\partial_\phi,
\end{eqnarray}
with dual
\beq
\fl\qquad
\omega^{\hat r}=-M(M_r \rmd r +M_\theta \rmd \theta), \quad
\omega^{\hat \theta}=\sqrt{g_{\theta\theta}}\rmd \theta, \quad
\omega^{\hat \phi}=\sqrt{g_{\phi\phi}}\rmd \phi.
\eeq
When expressed in terms of this frame, the spacetime metric (\ref{cmetu}) can be written as
\beq
g=-\omega^{\hat u}\otimes \omega^{\hat u}+ \omega^{\hat r}\otimes \omega^{\hat r}+\omega^{\hat \theta}\otimes \omega^{\hat \theta}+\omega^{\hat \phi}\otimes \omega^{\hat \phi}. 
\eeq

In this paper we present a novel study of the geodesic motion of photons on the equatorial plane of the exterior vacuum C metric, i.e. a gravitational background of a uniformly  accelerated static black hole.
It is well known, in fact, that these orbits can be integrated (in general and not only on the equatorial plane), due to the existence of the conformal Killing tensor (\ref{ckt}). 

We will show that the radial equation for equatorial  motion  can be put in exact correspondence with the case of equatorial geodesic motion of photons in the Schwarzschild background, a part from an energy shift of the form:
\beq
E_0 \quad ({\rm Schwarzschild})\quad \to \quad   \sqrt{E_0^2+A^2L_z^2} \quad ({\rm C metric})
\eeq
which evidentiates a coupling between the background acceleration $A$  and the (conserved) angular momentum ($L_z$) of photons.
Furthermore, photons starting their motion from the equatorial plane (with zero component of momentum along $\theta$: $P^\theta=0$) remain confined on it; the same property does not hold for massive particles which, instead,  are forced to escape from it  due to effects of the background acceleration itself.
Inspecting the massless limit of the C metric,  we will show that
this effect actually has its origin in Minkowski spacetime endowed with \lq\lq Bondi coordinate system," but it is lost if the Schwarzschild solution is considered, i.e. in the zero acceleration limit of the C metric.

We then discuss the possibility of experimentally identify the equatorial plane itself. We will succeed  in setting up a Fermi coordinate system for an observer O at rest at a fixed spatial point of the equatorial plane, giving also the transformation law between  Bondi-like spherical coordinates and Fermi coordinates explicitly (up to the second order). 
Finally, we suggest an experiment of light deflection as seen from the lab of the observer O which can be used to estimate the background acceleration parameter as well as the mass of the source.
We argue that these results can be applied to suitable astrophysical systems to set upper limits to a spacetime acceleration.

\section{Geodesics in $\{u, r, \theta, \phi\}$ coordinates}

The geodesic equations in $\{u,r,\theta,\phi\}$ coordinates are given by
\begin{eqnarray}
\label{geo1}
\fl
&\ddot u -\frac{H_r}{2}\dot u{}^2 +2Ars\dot u \dot \theta +\frac{rs^2}{G}\dot \theta{}^2
+rG \dot \phi{}^2=0 ,  \\
\label{geo2}
\fl
& \ddot \phi + \dot \phi \left[
\frac{2}{r}\dot r + \frac{G_\theta}{G}\dot \theta 
\right]=0 ,  \\
\label{geo3}
\fl
& \ddot r+\frac12 \left[ (H+A^2r^2G)H_r+\frac{AG}{s}H_\theta \right]\dot u{}^2
-\left[ r^3A^2s^2+\frac{Ar^2sG_\theta}{2G}+\frac{rs^2H}{G}\right]\dot \theta{}^2 \nonumber \\
\fl
&
-\left[ r^3A^2G^2+\frac{Ar^2GG_\theta}{2s}+rGH \right]\dot\phi{}^2+
\left[ H_r+2A^2rG\right]\dot u\dot r
\nonumber\\
\fl&
+\left[ H_\theta-2sAr(H+A^2r^2G)\right]\dot u\dot \theta =0
, \\
\label{geo4}
\fl
& \ddot \theta +\frac{G}{2r^2s^2}\left[Ar^2s H_r +H_\theta\right]\dot u{}^2
+\left[ \frac{c-Ars^2}{s}-\frac{G_\theta}{2G}\right]\dot \theta{}^2
-\frac{G}{2s^2}\left[G_\theta+2GArs \right]\dot \phi{}^2 \nonumber\\
\fl
&
+
\frac{2GA}{rs}\dot u \dot r
-2A^2rG\dot u\dot \theta +\frac2r 
\dot r \dot \theta =0\ ,
\end{eqnarray}
where a dot means differentiation with respect to an affine parameter $\lambda $ along the curve.

By using the Killing symmetries of the metric one gets the conserved quantities: the angular momentum $L_z$, associated to the Killing vector $\partial_\phi$, and the energy $E_0$, associated to the Killing vector $\partial_u$
\begin{eqnarray}
\label{geoEL}
\dot \phi =\frac{L_z}{r^2G}, \qquad
H \dot u+\dot r -r^2As 
\dot \theta =E_0 .
\end{eqnarray}
Furthermore, one has to specify if the particles are massive or null, i.e. one has the additional constraint
\beq
\label{massconstra}
H \dot u{}^2-2E_0 \dot u+\frac{r^2s^2}{G}
\dot \theta{}^2+\frac{L_z^2}{r^2G}=-\mu^2.
\eeq
For timelike (spacelike) geodesics, $\lambda$ can be identified with the proper time (spacelike curvilinear abscissa) and $\mu$ can be taken as $1$ ($i$).
Timelike geodesics has been firstly discussed by Pravda and Pravdov\'a \cite{pravdacqg}, by using different coordinate systems (symmetry-adapted coordinate systems and different from the present one too). 
They constructed an effective potential, whose properties allowed them to distinguish three different types of timelike geodesics, corresponding to particles (a) falling under the black hole horizon, (b) crossing the acceleration horizon and reaching future infinity, (c) spinning around the $z$-axis, co-accelerating with the black holes and then reaching future infinity. 
In the $(u,r,\theta, \phi)$ coordinate system the effective potential can be introduced rewriting the mass constraint
(\ref{massconstra}) as follows
\beq
\left( H\dot u -E_0 \right)^2+\frac{r^2s^2H}{G}\dot \theta{}^2=
E_0^2-V_{\rm (eff)}^2,
\eeq
with
\beq
V_{\rm (eff)}^2=H\left(\frac{L_z^2}{r^2G}+\mu^2\right).
\eeq

Null geodesics instead, have been all discussed by Kinnersley and Walker in 1970 \cite{kinwal}.
Following their results, the vector $P$  tangent to a null geodesic can be written as follows
\beq
\label{geonul}
\fl
P=\frac{1}{H}[E_0+R(r,\theta)]\partial_u -[R(r,\theta)+AV(\theta)]\partial_r -\frac{1}{r^2s}V(\theta)\partial_\theta+\frac{L_z}{r^2G}\partial_\phi ,
\eeq
where
\begin{eqnarray}
R(r,\theta)= \pm \sqrt{E_0^2-\frac{QH}{r^2}}, \qquad
V(\theta)=\pm \sqrt{QG-L_z^2}.
\end{eqnarray}

\subsection{Equatorial orbits: $\theta=\pi/2$}
Let us specialize our discussion  to the equatorial plane ($\theta=\pi/2$), considering the static spacetime portion lying in between the deformed Schwarzschild and Rindler horizons, located at
$$
r_S=\frac{-U+\sqrt{3}V}{A}, \qquad r_R=\frac{2U}{A},
$$ 
respectively (see Eqs.~(\ref{horS}) and (\ref{horR})). We denote the momentum of photons as $P^\mu=\rmd x^\mu/ \rmd \lambda$.
The geodesic equations in this case  ($\theta(\lambda)=\pi/2$) reduce to
\begin{eqnarray}
\label{geo1eq}
\fl\quad
&\ddot u-
\frac{H_r}{2}\dot u{}^2 
+r\dot\phi{}^2=0 , \qquad \ddot \phi + \frac{2}{r}\dot\phi 
\dot r =0 ,  \nonumber\\
\fl\quad
& \ddot r +\frac12 \left[ (H+A^2r^2)H_r+AH_\theta \right]\dot u{}^2 -r\left[H +r^2A^2 \right]\dot \phi{}^2
+
\left[ H_r+2A^2r\right]\dot u \dot r =0
, \nonumber\\
\fl\quad
& \frac{1}{2r^2}\left[Ar^2H_r +H_\theta\right]\dot u{}^2
-Ar \dot \phi{}^2 +
\frac{2A}{r}\dot u \dot r =0.
\end{eqnarray}
where we have used the $\theta \to \pi/2$ limit of $G$, $H$ and their derivatives
\beq
\fl
G=1,\quad G_\theta=0,\quad H=1-\frac{2m}{r}-A^2r^2,\quad H_r=\frac{2m}{r^2}-2rA^2, \quad H_\theta=2A(r-3m).
\eeq

The conserved quantities (\ref{geoEL}) and the mass constraint (\ref{massconstra}) become
\begin{eqnarray}
\label{geoLEeq}
\fl\qquad 
\dot \phi =\frac{L_z}{r^2}, \qquad
H \dot u +\dot r  =E_0, \qquad
H \dot u{}^2-2E_0 \dot u +\frac{L_z^2}{r^2}=-\mu^2 .
\end{eqnarray}
Eq.~(\ref{geo1eq})$_2$ is equivalent to Eq.~(\ref{geoLEeq})$_1$.
Eq.~(\ref{geo1eq})$_3$ can be rewritten as
\begin{eqnarray}
\fl
\ddot r + (H+A^2r^2)
\left[
\frac12 H_r\dot u{}^2 -r \dot \phi{}^2
\right]
+\frac{AH_\theta}{2} \dot u{}^2 
+
\left[ H_r+2A^2r\right]\dot u \dot r =0;
\end{eqnarray}
using Eq.~(\ref{geo1eq}),  one has
\begin{eqnarray}
\fl
\ddot r +\left[(H+A^2r^2)\ddot u 
+
\left( H_r+2A^2r\right)\dot u \dot r \right]+\frac{AH_\theta}{2} \dot u{}^2=0,
\end{eqnarray}
that is
\begin{eqnarray}
\frac{\rmd }{\rmd \lambda }\left[\dot r +(H+A^2r^2)\dot u \right]+
\frac{AH_\theta}{2} \dot u{}^2=0.
\end{eqnarray}
Using the Killing constraint (\ref{geoLEeq})$_2$ the latter equation becomes
\begin{eqnarray}
A\left[ 2r \dot r  \dot u  +r^2 \ddot u \right]
+
\frac{H_\theta}{2} \dot u{}^2=0.
\end{eqnarray}
Replacing $\dot r $ using again the Killing constraint (\ref{geoLEeq})$_2$ and $\ddot u $ using Eq.~(\ref{geo1eq}), we find 
\begin{eqnarray}
\label{compatib}
2rA E_0 \dot u -rAH \dot u{}^2- \frac{AL_z^2}{r}=0
\end{eqnarray}
which can be compatible only with  photons: $\mu=0$, as it follows by direct comparison with Eq.~(\ref{geoLEeq})$_3$.
Hence, the first important result that on the equatorial plane of the C metric only photons can be geodesic. Massive particles on this plane should necessary be accelerated.

Using the relation $(Ar^2H_r +H_\theta)=2ArH$, Eq.~(\ref{geo1eq})$_4$ can be rewritten as
\begin{eqnarray}
H\dot u{}^2
-r^2 \dot \phi{}^2 +
2\dot u \dot r =0,
\end{eqnarray}
which again coincides with Eq.~(\ref{geoLEeq})$_3$ with $\mu=0$.

Summarizing, the motion of geodesic photons on the equatorial plane of the C metric $(\theta=\pi/2)$ is described by
\begin{eqnarray}
\label{finali}
\fl\quad \dot u  &=& \frac1{H}\left[E_0-\dot r \right],\qquad 
\dot \phi =\frac{L_z}{r^2}, \nonumber \\
\fl\quad
\dot r &=& \pm \left( E_0^2 -\frac{HL_z^2}{r^2}\right)^{1/2}= \pm \left[ E_0^2+A^2L_z^2 -\frac{L_z^2}{r^2}\left(1-\frac{2m}{r}\right)\right]^{1/2}, 
\end{eqnarray}
in agreement with the solution of general null geodesic motion as described by Kinnersly and Walker (and  above rederived in Eq.~(\ref{geonul})).

Finally, it is convenient to rewrite the radial equation in terms of the impact parameter $b=L_z/E_0$ and an effective potential, so that
\beq
\fl\quad
\left( \frac{\rmd r}{\rmd \lambda}\right)= \pm L_z^2 \left[\frac{1}{b^2}-V^2_{\rm (eff)}\right]^{1/2}, \qquad
V^2_{\rm (eff)}=\frac{1}{r^2}\left(1-\frac{2m}{r}\right)-A^2\equiv \frac{H}{r^2} .
\eeq
The effective potential coincides with that of Schwarzschild modulo a costant shift, $-A^2$. Therefore, as in Schwarzschild 
the potential is peaked at $r=3m$ with a value 
$$V_{\rm (eff)}(3m)^2=\frac{1}{27m^2}-A^2\,.$$  

It is worth noticing that the radial equation (\ref{finali})$_2$ can be put itself in exact
 correspondence with the analogous radial equation for null geodesics of the equatorial plane of the Schwarzschild solution, by defining
 \beq
 \frac{1}{b_S^2}=\frac{1}{b^2}+A^2\qquad \to \qquad b_S=\frac{b}{\sqrt{1+A^2b^2}}.
 \eeq
Hence, as all possible  kind of equatorial null geodesics in the  Schwarzschild spacetime have been completely determined in terms of elliptic function \cite{chandra},  this problem here is also completely solved.
In fact, let us consider the orbit as parametrized by $\phi$, i.e.
\beq
\label{phipara}
\left( \frac{\rmd r}{\rmd \phi}\right)^2= r^4\, \left[\frac{1}{b_S^2}-\frac{1}{r^2}\left(1-\frac{2m}{r}\right)\right],
\eeq 
and introduce the new variable: $v(\phi)=2m/r(\phi)$. Eq.~(\ref{phipara}) thus becomes
\beq
\fl\quad
\left( \frac{\rmd v}{\rmd \phi}\right)^2=\left[\Omega^2-v^2\left(1-v \right)\right], \qquad \Omega=\frac{2m}{b_S}=\frac{2mb}{\sqrt{1+A^2b^2}},
\eeq 
and the solutions of this equation have been fully studied  in \cite{chandra} where all the details can be found.


\subsection{The equatorial plane in the massless (flat) limit of the  C metric}

In the case $m=0$ the C metric reduces to the Minkowski spacetime in Bondi coordinates
\begin{equation}
\label{cmetu_mink}
\fl
\rmd s^2= -[(1-Arc)^2-A^2r^2] \rmd u^2 - 2 \rmd u \rmd r + 2A r^2 s\,\rmd u \rmd \theta +r^2 \rmd \theta^2 +r^2 s^2 \rmd \phi^2  .
\end{equation} 
The map between Bondi and cartesian coordinates is easily obtained passing first to Rindler coordinates. In detail,
start from the form of the metric in standard cartesian coordinates $t,x,y,z$  
$$\rmd s^2=-\rmd t^2+\rmd x^2+\rmd y^2 +\rmd z^2\equiv \eta_{\mu\nu }\rmd x^\mu \rmd x^\nu .$$ 
Pass to Rindler coordinates $T,X,Y,Z$ using the map
\begin{eqnarray}
t&=& \left(\frac1A-Z\right)\sinh (AT), \nonumber \\
x&=&X, \nonumber \\
y&=&Y, \nonumber \\
z&=& -\left(\frac1A-Z\right)\cosh (AT),
\end{eqnarray}
so that the metric has the Rindler form
\beq
\label{Rstd}
\rmd s^2= -(1-AZ)^2 \rmd T^2+\rmd X^2+\rmd Y^2+\rmd Z^2\ ,
\eeq
with the limitations $X\in (-\infty, +\infty)$,  $Y\in (-\infty, +\infty)$, $Z \in (-\infty, +1/A)$.
Finally pass to Bondi coordinates $u,r,\theta,\phi$ with
\begin{eqnarray}
\label{transf_flat}
T&=&-u+\frac{1}{2A} \ln \left[\frac{1-Ar(1+\cos \theta)}{1+Ar(1-\cos \theta)}\right], \nonumber \\
X&=&r \sin \theta \cos \phi, \nonumber \\
Y&=&r \sin \theta \sin \phi, \nonumber \\
Z&=&\frac1A-\sqrt{\left(\frac1A-r\cos \theta\right)^2-r^2},
\end{eqnarray}
which brings the metric (\ref{Rstd}) in the Bondi form (\ref{cmetu_mink}).
The usual interpretation of $r$ as a radial coordinate gives the limitation 
\beq
r<\frac{1}{A(1+c)};
\eeq
For the other coordinates instead the limitations are the usual: $u\in (-\infty , \infty)$, $\theta\in (0,\pi)$, $\phi \in (0,2\pi)$.

We are now ready to determine the shape of the hypersurface $\theta=\pi/2$, once represented in cartesian coordinates as well as to study geodesics confined on this surface, having in mind what happens in the case of the C metric.
It results
\beq
\cot \theta =\frac{A}{2\sqrt{x^2+y^2}}\left( \frac{1}{A^2}+t^2-x^2-y^2-z^2\right);
\eeq
hence $\theta=\pi/2$ corresponds to the hyperboloid:
\beq
\label{hyp}
\frac{1}{A^2}+t^2-x^2-y^2-z^2=0.
\eeq
Moreover, the geodesics for Minkowski spacetime in cartesian coordinates are straight lines:
\beq
\fl
t=t_0 + \lambda E_0, \quad x=x_0+\lambda P_x, \quad y=y_0+\lambda P_y, \quad z=z_0+\lambda P_z.
\eeq
In order to remain confined on the hyperboloid  (\ref{hyp}), corresponding to $\theta=\pi/2$, the equation 
\beq
\fl
\frac{1}{A^2}+(t_0 + \lambda E_0)^2-(x_0+\lambda P_x)^2-(y_0+\lambda P_y)^2-(z_0+\lambda P_z)^2=0
\eeq
must be satisfyied for any $\lambda$.
This gives the following relations:
\begin{eqnarray}
& \frac{1}{A^2}+t_0^2-x_0^2-y_0^2-z_0^2=0, \nonumber \\
& E_0^2=P_x^2+P_y^2+P_z^2, \nonumber \\
& t_0E_0-x_0P_x-y_0P_y-z_0P_z=0.
\end{eqnarray}
The interpretation is easy: 1) the starting point $P_0=(t_0,x_0,y_0,z_0)$ must belong to the hyperboloid itself. Hence the parametrization
\begin{eqnarray*}
& t_0=\frac{1}{A}\sinh \psi_1, \quad x_0=\frac{1}{A}\cosh \psi_1\sin\psi_2\cos\psi_3, \nonumber \\
& y_0=\frac{1}{A}\cosh \psi_1\sin\psi_2\sin\psi_3 ,\quad z_0=\frac{1}{A}\cosh \psi_1\cos\psi_2, 
\end{eqnarray*} 
can be useful; 2) the geodesics must be null; 3) the momentum must be orthogonal to the normal to the hyperboloid in the point $P_0$, i.e. tangent to the hyperboloid.
As all the null vectors emanating from $P_0$ generate a null cone, geodesics on the equatorial plane are defined by those null vectors obtained intersecting the null cone and the hyperboloid itself. Explicitly, one rewrites the spatial components of the momentum as follows
$$P_x=E_0\sin \alpha \cos \beta, \quad P_y=E_0\sin \alpha \sin \beta, \quad P_z=E_0\cos \alpha $$
and looks for the solutions $\alpha$ and $\beta$ of the equation:
\begin{eqnarray}
& t_0-\sin \alpha (x_0\cos \beta-y_0 \sin \beta)-z_0\cos \alpha=0,
\end{eqnarray}
which are in general $\infty^1$.
It is worth to note that when studying instead the Schwarzschild limit of the C metric ($A=0$),  both photons and massive particles can be confined on the equatorial plane $\theta=\pi/2$. The equatorial plane of the C metric seems thus to inherit the corresponding property of Minkowski spacetime in Bondi coordinates.

\section{\lq\lq Lab" on the equatorial plane of the C metric: Fermi coordinates}

In the previous section we have shown that a peculiar role is played by the equatorial plane of the C metric for what concerns photons. We have also mentioned that this  property still survives for the equatorial plane of the Minkowski solution written in Bondi coordinates (the massless limit of the C metric) but it is lost for the zero acceleration limit of the C metric (i.e. the Schwarzschild solution): here, in fact,  condition (\ref{compatib}), being proportional to $A$, is automatically satisfied for $A=0$ and it does not give restrictions to the mass of the particles. 

Therefore, it is quite natural to study the possibility to put an observer with his own \lq\lq Lab" on the equatorial plane of the  C metric and consider the \lq\lq observer proper frame", naturally defined in terms of Fermi coordinates adapted to his world line. 

The coordinate transformation from the Bondi-like spherical coordinates $\{u,r,\theta,\phi \}$ to the Fermi coordinates $\{T,X,Y,Z\}$ adapted to  $e_{\hat u}$  can be obtained following the procedure described in \cite{bgj_FC}: 
\begin{eqnarray}
\label{FCCM}
\fl\quad u&=& u_0+\frac{1}{\sqrt{H_0}}(T-X)+\frac{AXY}{\sqrt{H_0}}
+\frac{1}{2r_0}\left[\left(1-\frac{r_0-3m}{r_0H_0}\right)X^2-Y^2-Z^2\right],\nonumber \\
\fl\quad r&=& r_0+\sqrt{H_0}X+Ar_0Y+\frac{m}{2r_0^2}X^2+\frac{r_0-2m}{2r_0^2}(Y^2+Z^2),\nonumber \\
\fl\quad \theta&=&\frac{\pi}{2} +\frac{Y}{r_0}\left(1 -\frac{\sqrt{H_0}}{r_0}X\right)+\frac{A}{2r_0}(X^2-Y^2+Z^2), \nonumber \\
\fl\quad \phi &=&\phi_0 +\frac{Z}{r_0}\left( 1-AY-\frac{\sqrt{H_0}}{r_0}X\right) ,
\end{eqnarray}
where $H_0=1-2m/r_0-A^2r_0^2$.
This coordinate transformation maps the Bondi-like form (\ref{cmetu}) of the C metric  into the (flat) \lq\lq Rindler" form, up to the second order in the coordinates $T, X,Y,Z$:
\beq
\fl\quad
\label{FCform}
\rmd s^2= -(1-2G_1X-2G_2Y)\, \rmd T^2 +\rmd X^2 +\rmd Y^2 +\rmd Z^2 + O(3),
\eeq
with
\beq
 G_1= -\frac{\sqrt{H_0}}{r_0}+\frac{r_0-3m}{r_0^2\sqrt{H_0}}, \qquad G_2=-A, 
\eeq
the coordinate components of the observer's acceleration.
To give a geometrical interpretation of Eq.~(\ref{FCCM}), we find convenient to introduce the following linear combinations:
\begin{eqnarray}
\Delta u &=&M (u-u_0)-MM_r (r-r_0)-MM_\theta (\theta-\theta_0), \nonumber \\
\Delta r &=&-MM_r (r-r_0)-MM_\theta (\theta-\theta_0) , \nonumber \\
\Delta \theta &=&\sqrt{g_{\theta\theta}}(\theta-\theta_0) , \nonumber \\
\Delta \phi &=&\sqrt{g_{\phi\phi}}(\phi-\phi_0) , 
\end{eqnarray}
so that
\begin{eqnarray}
\Delta u &=& T + \kappa (r, e_{\hat u})^{\hat r} X^2 + 2AXY  - \tilde \kappa (\theta, e_{\hat u})^{\hat \theta} Y^2, \nonumber \\
\Delta r &=&  X+AXY +\frac12 \kappa (r, e_{\hat u})^{\hat r} X^2 -\frac12 \kappa (\phi, e_{\hat u})^{\hat r} (Y^2+Z^2), \nonumber \\
\Delta \theta &=& Y+\frac{A}{2} (X^2-Y^2+Z^2)+  \kappa (\phi, e_{\hat u})^{\hat r} XY , \nonumber \\
\Delta \phi &=& Z -AYZ +\kappa (\phi, e_{\hat u})^{\hat r} XZ , 
\end{eqnarray}
where we have introduced the curvatures associated to the coordinate lines: 
$\kappa (x^a, e_{\hat u})_{\hat c}=-e_{\hat c}\ln \sqrt{\gamma_{aa}}$:
\begin{eqnarray}
 \kappa (r, e_{\hat u})^{\hat r}&=\frac{m-A^2r_0^3}{r_0^2\sqrt{H_0}}, \qquad & \kappa (r, e_{\hat u})^{\hat \theta} =A,\nonumber \\
 \kappa (\theta, e_{\hat u})^{\hat r}&=-\frac{(r_0-2m)^2-mA^2r_0^3}{r_0^2(r_0-2m)\sqrt{H_0}}  , \qquad & \kappa (\theta, e_{\hat u})^{\hat \theta}=-A,
\nonumber \\
 \kappa (\phi, e_{\hat u})^{\hat r}&=-\frac{\sqrt{H_0}}{r_0},  \qquad & \kappa (\phi, e_{\hat u})^{\hat \theta}=-A,
\end{eqnarray}
together with another useful quantity 
\beq
{\tilde \kappa} (\theta, e_{\hat u})^{\hat \theta}\equiv  -\frac{1}{r_0}e_{\hat \theta}(MM_\theta)=-\frac{A^2r_0}{\sqrt{H_0}},
\eeq
which seems to play the same role as a curvature.
The two limits $m=0$ and $A=0$ of these relations immediately follow.

In the new form (\ref{FCform}) of the C metric the solution for the geodesic equations is
\begin{eqnarray}
\label{geosfermi}
T(\lambda)&=&C\lambda^2 (G_1X_1+G_2Y_1)+C\lambda (1+2G_1X_0+2G_2Y_0)+T_0\ , \nonumber\\
X(\lambda)&=&\frac{C^2G_1}2\lambda^2+X_1\lambda+X_0\ , \nonumber\\
Y(\lambda)&=&\frac{C^2G_2}2\lambda^2+Y_1\lambda+Y_0 , \nonumber\\
Z(\lambda)&=&Z_1\lambda+Z_0,
\end{eqnarray}
where $C=\sqrt{(\mu^2+X_1^2+Y_1^2+Z_1^2)(1-2G_1X_0-2G_2Y_0)}$, and $\mu=0$ for photons ($\mu=1$ for massive particles if proper time parametrization of the orbits is chosen).

Let us confine our attention to the motion in the $X-Y$ plane ($Z_0=0=Z_1$) of a particle emitted at the origin ($X_0=0=Y_0$).
By eliminating the affine parameter from Eqs.~(\ref{geosfermi})$_2$ and (\ref{geosfermi})$_4$, i.e.
\beq
\lambda=\frac{G_1X-G_2Y}{G_1X_1-G_2Y_1}\ ,
\eeq 
and then substituting for instance into Eq.~(\ref{geosfermi})$_3$, we obtain the following implicit form for the parametric equation of the trajectory:
\beq
0=\frac{G_1X-G_2Y}{G_1X_1-G_2Y_1}\left[\frac{G_1X-G_2Y}{G_1X_1-G_2Y_1}\frac{C^2G_1}{2}+Y_1\right]-Y\ .
\eeq
The vertical deflection of a particle horizontally emitted ($Y_1=0$) at the origin is given by the coordinate
\beq
Y=\frac{C^2}{2X_1^2}G_1(\delta X)^2\ , \qquad C^2=\mu^2+X_1^2\ ,
\eeq
where $\delta X$ denotes a small displacement along the horizontal direction.
From the previous equation, the light deflection turns out to be 
\beq
Y=\frac12G_1(\delta X)^2\ , 
\eeq  
as expected.
Analogously, the horizontal deflection of a particle vertically emitted ($X_1=0$) at the origin is given by the coordinate
\beq
X=\frac{C^2}{2Y_1^2}G_2(\delta Y)^2\ , \qquad C^2=\mu^2+Y_1^2\ ,
\eeq
where $\delta Y$ denotes a small displacement along the vertical direction;
the corresponding light deflection is 
\beq
X=\frac12G_2(\delta Y)^2\ . 
\eeq  
Being $G_2=-A$, it is clear that experiments of light deflection in labs, placed on the equatorial plane of a C metric solution, may be useful to determine the background acceleration parameter $A$. 
Moreover, the combined measurements of vertical and horizontal deflections would allow in principle to estimate both the mass and the acceleration parameters of the source.
However, the effective feasibility of such experiments is beyond the scope of the present paper.

\section{Concluding remarks}

We have presented a novel discussion of geodesics in the vacuum C metric, using Bondi-like spherical coordinates, which evidentiates a special role played by the equatorial plane. We have first shown that the exact integrability of photon orbits lying (with their spatial trajectory)  on the 
equatorial plane can be put in exact correspondence with the analogous orbit of  photons on the equatorial plane of the Schwarzschild spacetime, apart from an energy shift involving the spacetime acceleration parameter.
Furthermore, we have shown that photons starting their motion from this hypersurface and with initial momentum contained in the tangent hyperplane, remain confined on it, whereas the same is not true for massive particles.

Finally, we have given the explict map between Bondi-like spherical coordinates and Fermi coordinates (up to the second order) for the world line of an observer at rest at a fixed spatial point of the equatorial plane of the C metric, possibly a building-block for the construction of any lab and measurement process. 
We have then suggested a possible procedure to estimate both the mass and acceleration parameters of the background source by measuring  light deflection.
We expect that these results might be applied to accelerated sources of astrophysical interest.

\end{document}